\documentclass[lettersize,journal]{IEEEtran}
\usepackage{amsmath,amsfonts}
\usepackage{algorithmic}
\usepackage{algorithm}
\usepackage{array}
\usepackage[caption=false,font=normalsize,labelfont=sf,textfont=sf]{subfig}
\usepackage{textcomp}
\usepackage{stfloats}
\usepackage{url}
\usepackage{verbatim}
\usepackage{graphicx}
\usepackage{cite}
\usepackage{multirow}
\usepackage{xcolor}
\begin{document}

\title{Power-balanced Memristive Cryptographic Implementation Against Side Channel Attacks}

\author{
    \IEEEauthorblockN{Ziang Chen$^{1,2}$, Li-Wei Chen$^{3}$, Xianyue Zhao$^{1,2}$, Kefeng Li$^{1,2}$, Heidemarie Schmidt$^{1,2}$, Ilia Polian$^{3}$, Nan Du$^{1,2,*}$}
    
    \IEEEauthorblockA{$^1$ Institute for Solid State Physics, Friedrich Schiller University Jena, Jena, Germany}
    
    \IEEEauthorblockA{$^2$ Department of Quantum Detection, Leibniz Institute of Photonic Technology, Jena, Germany}
    
    \IEEEauthorblockA{$^3$ Institute of Computer Science and Computer Engineering, University of Stuttgart, Stuttgart, Germany}
    
    \thanks{Corresponding author: Dr. Nan Du, nan.du@uni-jena.de}
}



\markboth{IEEE Transactions on Nanotechnology}%
{Chen \MakeLowercase{\textit{et al.}}: Power-Balanced Memristive Cryptography}


\maketitle

\begin{abstract}
Memristors, as emerging nano-devices, offer promising performance and exhibit rich electrical dynamic behavior. Having already found success in applications such as neuromorphic and in-memory computing, researchers are now exploring their potential for cryptographic implementations. In this study, we present a novel power-balanced hiding strategy utilizing memristor groups to conceal power consumption in cryptographic logic circuits. Our approach ensures consistent power costs of all 16 logic gates in Complementary-Resistive-Switching-with-Reading (CRS-R) logic family during writing and reading cycles regardless of Logic Input Variable (LIV) values. By constructing hiding groups, we enable an effective power balance in each gate hiding group. Furthermore, experimental validation of our strategy includes the implementation of a cryptographic construction, xor4SBox, using NOR gates. The circuit construction without the hiding strategy and with the hiding strategy undergo T-test analysis, confirming the significant improvement achieved with our approach. Our work presents a substantial advancement in power-balanced hiding methods, offering enhanced security and efficiency in logic circuits.
\end{abstract}

\begin{IEEEkeywords}
Memristive technology, CRS-R logic family, Boolean logic gate, Side-channel attacks, Power-balanced hiding strategy, Cryptographic circuits.
\end{IEEEkeywords}

\section{Introduction}
\label{sec:Introduction}
\IEEEPARstart{M}{emristors} exhibit remarkable properties that position them as a highly promising technology with a wide range of applications, including non-volatile memories and neuromorphic computing \cite{kumar2022dynamical}. In particular, their ability to enable in-memory computing (IMC) architectures makes them highly attractive for cryptographic applications \cite{du2021low, Polian2023}. This is due to the advantage of keeping sensitive information, such as secret keys, unencrypted plaintexts, or internal cipher states, within the same memory and compute modules, eliminating the need for data transfer between separate components.

Memristors are popular for serving as security anchors in implementation of cryptographic primitives, such as physical unclonable functions (PUFs) and true random number generators (TRNGs), and cryptographic logic functions. PUFs utilize memristive device variations, such as cycle-to-cycle (C2C) and device-to-device (D2D) variations, to create unique and unpredictable behavior while suppressing unwanted variations. Recent memristive PUF designs exploit write time \cite{RMYW:13,KKS:13}, readout current \cite{GCLY:16}, and readout resistance differences \cite{JSV+:21}. Resilient memristive PUF constructions, including erasable PUFs \cite{GJK+18} and concealable PUFs \cite{GLP+:22}, enhance security and protect sensitive information. Memristor blocks can function as standalone RNGs or integrated functions like MemHash \cite{AK:17}. Memristive TRNGs utilize sources such as random telegraph noise \cite{HST+:12}, voltage variations \cite{BAWI:15,BAC+:16}, and readout current fluctuation randomness \cite{WKO+:16}. However, the evaluation procedures of these TRNGs often lack a comprehensive understanding of the noise source, falling short of cryptography standards. 

For cryptographic logic functions, various memristive logic families have been proposed \cite{DU2021186, PVV:14,BSK:10,KBL:14,KWS:12,YSL:14}, but experimental demonstrations of typical constructions face challenges. It is advisable to use well-established cryptographic function like the Advanced Encryption Standard (AES) block cipher, as modified designs can compromise security and interoperability \cite{LCL+:22}. Physical attack resistance in memristive logic requires further investigation, especially concerning the effects of variability on input-dependency \cite{CLP+:13}. Cryptographic modules based on memristive logic highlight vulnerabilities and the need for tailored security measures, including side-channel analysis resistance and the development of memristive-oriented power estimation models \cite{CLP+:13}.

Despite extensive research on using memristors for implementing cryptographic primitives, cryptographic logic functions, including the development of side-channel attacks (SCAs) that exploit the physical characteristics of memristive cryptographic implementations, such as power consumption, electromagnetic radiation, or timing, to uncover secret information like cryptographic keys, there have been relatively few works on studying countermeasures against these attacks, addressing this gap is crucial for ensuring the security and resilience of memristor-based cryptographic systems. 

Among different countermeasure techniques, one crucial class of countermeasures is hiding, which aims to minimize or eliminate the correlation between physical characteristics and secret information. These techniques make it more challenging for attackers to extract meaningful data from side-channel measurements. However, in the context of memristor technology, the exploration of hiding techniques for concealing the power of memristive cryptographic circuitry against SCAs remains largely unexplored.

In CMOS technology, Sense Amplifier-Based Logic (SABL) proposed by Tiri et al. \cite{tiri2002dynamic} served as an initial gate-level hiding technique. Operating under a Dual Precharge Logic (DPL) style, SABL ensured independent output switching regardless of input values, generating transitions by computing the output signal and its complement. However, its drawbacks included a full-custom logic style and the need for differential routing to maintain uniform load capacitance for complemented signals. 
The complexity of full-custom logic styles, which are unsuitable for conventional digital design flows and field programmable gate array (FPGA) implementations, led to the development of alternative designs by Tiri \cite{tiri2004logic}. This included Simple Dynamic Differential Logic (SDDL), Wave Dynamic Differential Logic (WDDL), and Divided Wave Dynamic Differential Logic (DWDDL). While SDDL, employing standard gates with differential inputs and outputs, had limitations in secure implementations, WDDL improved upon it by halving the area through the propagation of the precharge signal as a wave through combinational logic. DWDDL implemented WDDL with two distinct dual parts, achieving differential implementation with performance, security, and cost levels akin to WDDL.

In comparison to CMOS technology, memristors offer the advantage of eliminating the precharge process, significantly simplifying circuit design when used as an alternative for implementing logic gates in hiding techniques. Moreover, memristors accomplish logic gate functionality through their resistance state changes, enabling reusability. Consequently, memristor-based implementations in hiding techniques eliminate the need for both differential routing to balance wiring capacitances and other complex layout rules, substantially streamlining chip layout design.
Therefore, in this study, our focus is on developing countermeasure techniques specifically designed to conceal power-related information in memristive cryptographic circuitry, providing novel insights into this important area.

In our previous work \cite{you2014exploiting}, we introduced the BiFeTiO$_3$/BiFeO$_3$ (BiBFO) memristive device, which exhibits a single complementary bipolar resistive switching behavior capable of realizing all 16 Boolean logic functions in three logic cycles. Building upon this, in this study, we demonstrate the implementation of cryptographic logic functions using the CRS-R logic family of BiBFO memristive devices. Furthermore, we investigate a power-balanced based hiding technique for memristor cryptographic circuitry, leveraging the unique properties of the BiBFO memristive device. 

The paper has been structured as follows: Sec. \ref{sec:BiBFO} introduces the BiBFO memristor and elucidates its complementary bipolar resistive switching dynamics. The utilization of BiBFO memristors for implementing 16 Boolean logic gates is explored, and the development of our power-balanced hiding technique by employing BiBFO memristors is presented in Sec. \ref{sec:hidingTec}. Furthermore, the experimental validation of the hiding technique is performed on a xorSBox using T-test analysis in Sec. \ref{sec:Valid.}. Sec. \ref{sec:Conclu.} concludes the article.

\section{Self-rectifying complementary memristive devices}
\label{sec:BiBFO}

\begin{figure*}[htb]
  \includegraphics[width=\linewidth]{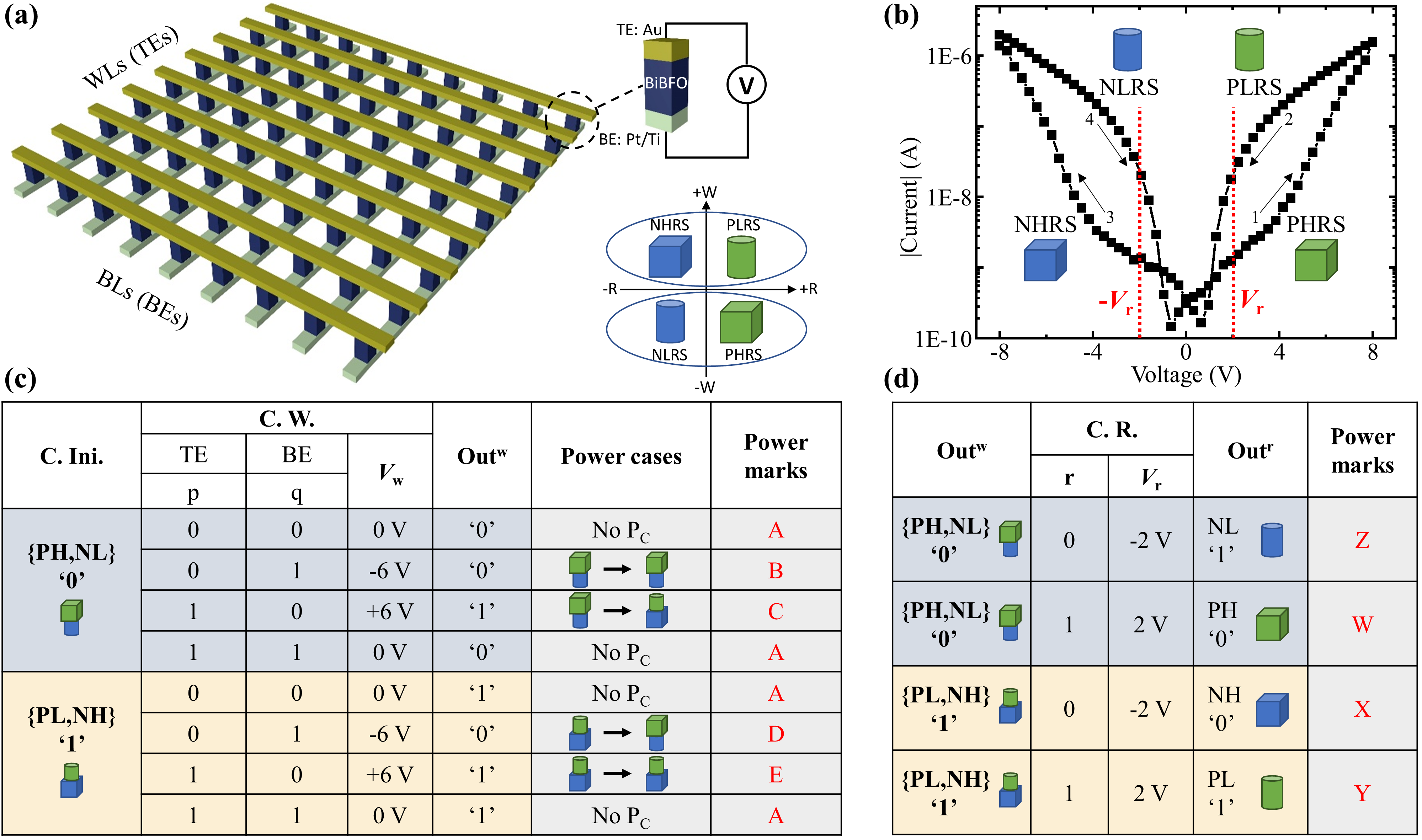}
  \caption{(a) Crossbar topology based on BiBFO memristors. Insets demonstrate the structure of BiBFO memristor and the correlation between resistance states of BiBFO memristor and writing/reading biases. (b) BiBFO memristors exhibit complementary bipolar $I-V$ characteristics when subjected to a triangle-shaped ramping bias. Truth tables and power consumption analysis for the writing and reading cycles of BiBFO memristors are presented in (c) and (d), respectively.}
  \label{fig:IV}
\end{figure*}

\begin{figure*}[htb]
  \includegraphics[width=\linewidth]{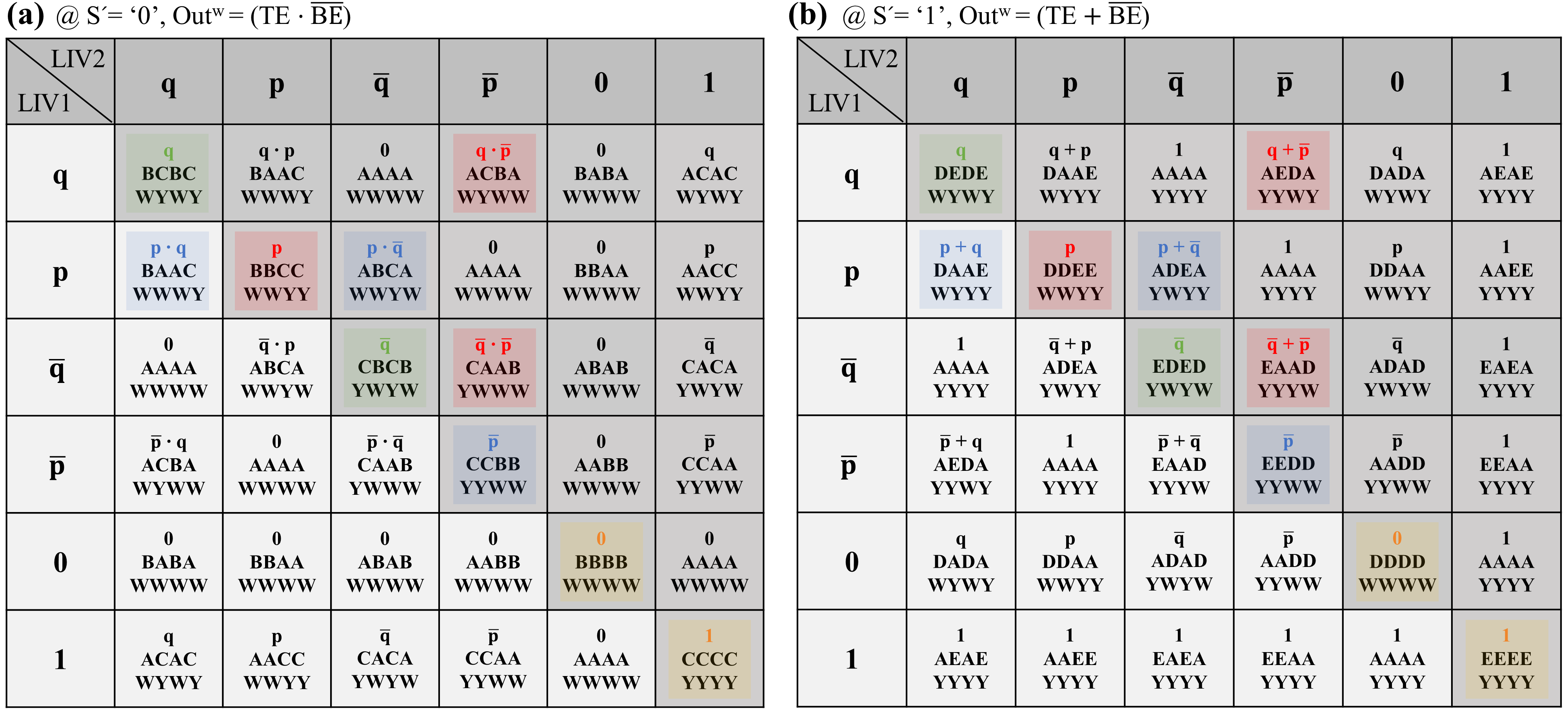}
  \caption{An overview of the power hiding strategy for achieving 14 gate types except XOR and XNOR by employing BiBFO memristors, with a single writing cycle and two different initialization states: (a) S' = '0' and (b) S' = '1'.}
  \label{fig:14gates}
\end{figure*}

The self-rectifying complementary BiBFO memristor, as illustrated in the inset of Fig. \ref{fig:IV}a, comprises  BiFeTiO$_3$ (BFTO)/BiFeO$_3$ (BFO) thin films sandwiched between an Au top electrode (TE) and a Pt/Ti bottom electrode (BE) on a SiO$_2$/Si substrate. For fabricating the BiBFO memristor, the polycrystalline BiFeTiO$_3$ (BFTO) and BiFeO$_3$ (BFO) thin films (nominal thickness of 550 nm) with perovskite structure (R3c space group) are deposited sequentially on Pt(150 nm)/Ti(50 nm)/SiO$_2$/Si substrate by using pulsed-laser deposition (PLD). The deposition process is conducted at PLD ambient temperature of 650 ~$^\text{o}$C, laser energy density of 2.1 J/cm$^2$, oxygen pressure of 1.3E-2 mTorr, and repetition frequency of 10 Hz. Upon deposition, the Au TE with size of 4.52E-2 mm$^2$ is sputtered on BFO thin film at room temperature using a metal shadow mask. 

The schematics of fabricated BiBFO memristor are demonstrated in the inset of Fig. \ref{fig:IV}a. By applying a triangle-shaped ramping bias with an amplitude of 8 V and a bias direction of 0~V $\rightarrow$ 8~V $\rightarrow$ -8~V $\rightarrow$ 0~V, the experimental current-voltage ($I-V$) characteristics of the BiBFO memristor are recorded and depicted in Fig. \ref{fig:IV}b. These characteristics exhibit the distinctive bipolar switching dynamics in the BiBFO cell, with the currents flowing through the memristive devices following the sequence of 1 $\rightarrow$ 2 $\rightarrow$ 3 $\rightarrow$ 4.

As shown in Fig. \ref{fig:IV}b, the BiBFO memristive device exhibits switching behavior between the low resistance state (LRS) and high resistance state (HRS) by applying writing pulses with opposite polarities. During the SET pulse, a positive voltage ($V_\text{w}$ = 6 V) is applied to the TE while the BE is grounded, resulting in the BiBFO memristor being in the LRS in the positive bias range (PLRS) and HRS in the negative bias range (NHRS). Conversely, during the RESET pulse, a negative voltage ($V_\text{w}$ = -6 V) is applied to the TE with the BE grounded, leading to the HRS in the positive bias range (PHRS) and LRS in the negative bias range (NLRS). The resistance states, represented by the PLRS (green cylinder) and PHRS (green cube), can be determined at a small reading bias $V_\text{r}$ of 2.5 V, while the NLRS (blue cylinder) and NHRS (blue cube) can be determined at a reading bias $V_\text{r}$ of -2.5 V. Therefore, the operation of a BiBFO memristor typically involves a sequential process consisting of three cycles: the initialization cycle (C. Ini.), the writing cycle (C. W.), and the reading cycle (C. R.).

To establish the correlation between resistance states and the polarities of writing and reading biases, we employ a 2D coordinate system where the x-axis represents the reading bias and the y-axis represents the writing bias. Four distinct resistance states, along with their corresponding symbols, are depicted in this coordinate system. Upon applying a positive writing voltage $V_\text{w}$, the BiBFO memristor transitions to the {PLRS, NHRS} ({PL, NH}) state, as illustrated by a combination of a green cylinder and a blue cube in the inset of Fig. \ref{fig:IV}a. Conversely, when a negative writing voltage $V_\text{w}$ is applied, the memristor assumes the {PHRS, NLRS} ({PH, NL}) state, as represented by a combination of a green cube and a blue cylinder. By employing positive and negative reading biases, we can achieve different resistance states, where distinct polarities of the reading bias result in diverse outcomes while maintaining a consistent writing bias.

\begin{figure*}[htb]
  \includegraphics[width=1\linewidth]{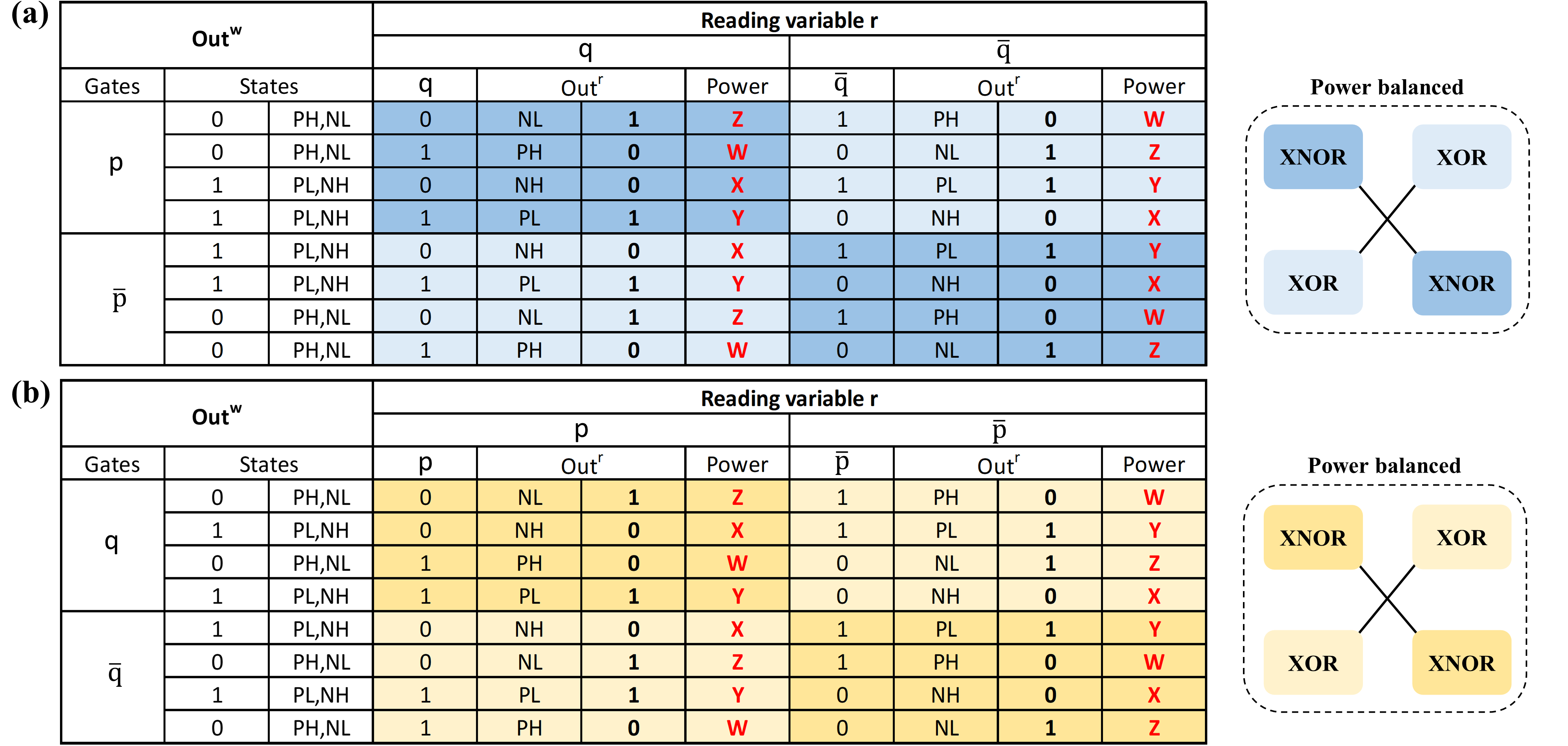}
  \caption{Configurations and power concealment in XOR and XNOR gate types by unitizing BiBFO memristors with reading cycles. (a) Hiding group of XOR and XNOR gate with $\text{Out}^\text{w}$ = p and $\overline{\text{p}}$ and with r = q and $\overline{\text{p}}$. (b) Hiding group of XOR and XNOR gate with $\text{Out}^\text{w}$ = q and $\overline{\text{q}}$ and with r = p and $\overline{\text{p}}$.}
  \label{fig:XOR}
\end{figure*}

The output states in the writing cycle ($\text{Out}^\text{w}$) of the BiBFO memristor are determined by the voltage difference (i.e., $V_\text{w}$) between the TE and BE, as well as the initialization state S' of the BiBFO memristor. The changes in the resistance state of the BiBFO memristor during the writing cycle can be effectively understood by referring to a truth table, which provides a comprehensive overview of the power consumption classifications for each specific case. This visualization of the dynamic resistance state changes and power consumption classifications is depicted in Fig. \ref{fig:IV}c.

In our framework, we have assigned the state {PH, NL} as '0', represented by a symbol composed of a green cube and a blue cylinder, and the state {PL, NH} as '1', represented by a symbol composed of a green cylinder and a blue cube. The writing bias is determined by the potential levels at the TE and BE, which are influenced by the logic variables p and q (where 1 represents high potential and 0 represents low potential). It is important to note that when TE and BE have the same potential (e.g., TE = 0 and BE = 0, or TE = 1 and BE = 1), there is no voltage difference across the device (i.e., $V_\text{w}$ = 0 V), resulting in an unchanged state and no power consumption ($\text{P}_\text{C}$). 

When TE is set to 0 and BE is set to 1, the corresponding $V_\text{w}$ is -6 V. In this case, regardless of the initialization resistance state S', the BiBFO memristor consistently resets to the {PH, NL} ('0') state. Conversely, when TE is set to 1 and BE is set to 0, the applied $V_\text{w}$ of 6 V reliably sets the BiBFO memristor to the {PL, NH} ('1') state.
Based on the variations in writing voltages $V_\text{w}$ and switching dynamics, the observed power consumption can be categorized into five distinct classes with corresponding power marks: A, B, C, D, and E. These categories enable comprehensive analysis and interpretation. Power mark A represents the absence of power dissipation, indicating no power consumption. Power marks B and D indicate power consumption resulting from the application of a -6 V $V_\text{w}$. Specifically, mark B signifies the persistence of the initialization resistance state {PH, NL}, while mark D represents the transformation from the initialization resistance state {PL, NH} to {PH, NL}. Similarly, power marks C and E capture power consumption during the utilization of a +6 V writing voltage. Mark E characterizes the sustained presence of the initialization resistance state {PL, NH}, while mark C signifies the transition from the initialization resistance state {PH, NL} to {PL, NH}.

The reading operation, following the writing cycle, plays a crucial role in realizing Boolean logic gates using the BiBFO memristor. The truth table and associated power consumption of the reading cycle are illustrated in Fig. \ref{fig:IV}d. During the reading cycle, the output states ($\text{Out}^\text{r}$) are obtained by reading the $\text{Out}^\text{w}$ using a specific reading bias ($V_\text{r}$). When the reading variable r is 0, a reading voltage $V_\text{r}$ of -2.5 V is applied to retrieve the resistance state from the negative voltage range. For example, if the $\text{Out}^\text{w}$ is {PH, NL}/{PL, NH}, the reading process will reveal the NL/NH state with $V_\text{r}$ = -2.5 V. 

Conversely, when r is 1, a reading voltage $V_\text{r}$ of 2.5 V is used to access the resistance state in the positive voltage range. Following the same example, the reading operation will uncover the PH/PL state with $V_\text{r}$ = 2.5 V. In our reading cycle framework, we assign logic '1' to the low resistance states of the $\text{Out}^\text{r}$, namely PL and NL, while the high resistance states, PH and NH, are designated as logic '0'. With a reading voltage amplitude of 2.5 V, the power consumption during the reading operation can be categorized into four distinct cases based on the observed resistance states of the $\text{Out}^\text{r}$, specifically marked as NL: Z, PH: W, NH: X, and PL: Y.

\section{Power balanced hiding technique in BiBFO}
\label{sec:hidingTec}

Using a BiBFO memristor with a single writing cycle and no reading cycle, it is possible to realize 14 gate types except the XOR and XNOR. Fig. \ref{fig:14gates} provides an overview of the power hiding strategy for achieving these 14 gates by employing BiBFO and a single writing cycle. In Fig. \ref{fig:14gates}, it should be noted that the reading variable r is set as 1. Importantly, regardless of whether a reading cycle is involved or not, the logic output variable after the writing cycle remains unchanged, represented as $\text{Out}^\text{r}$ = $\text{Out}^\text{w}$, when r is equal to 1. In Fig. \ref{fig:14gates}a, the TE is assigned the logic input variable 1 (LIV1), while the BE is assigned the negation of logic input variable 2 (LIV2), enabling the realization of the corresponding logic gate at the intersection of LIV1 and LIV2. Fig. \ref{fig:14gates} illustrates six LIV1/LIV2 possibilities: q, p, $\overline{\text{q}}$, $\overline{\text{p}}$, 0, and 1. For example, to achieve the gate $\overline{\text{q}}$ $\cdot$ $\overline{\text{p}}$, LIV1 = $\overline{\text{q}}$ is applied to the TE and $\overline{\text{LIV2}}$ = p is applied to the BE, following the provided equations on tables in Fig. \ref{fig:14gates}. The power consumption during the writing and reading cycles for implementing the $\overline{\text{q}}$ $\cdot$ $\overline{\text{p}}$ gate is indicated in Fig. \ref{fig:14gates}. The power marks for the writing cycle in 4 input cases (CAAB in the first row) and the reading cycle (YWWW in the second row) are annotated accordingly.
\begin{figure*}[htb]
  \includegraphics[width=1\linewidth]{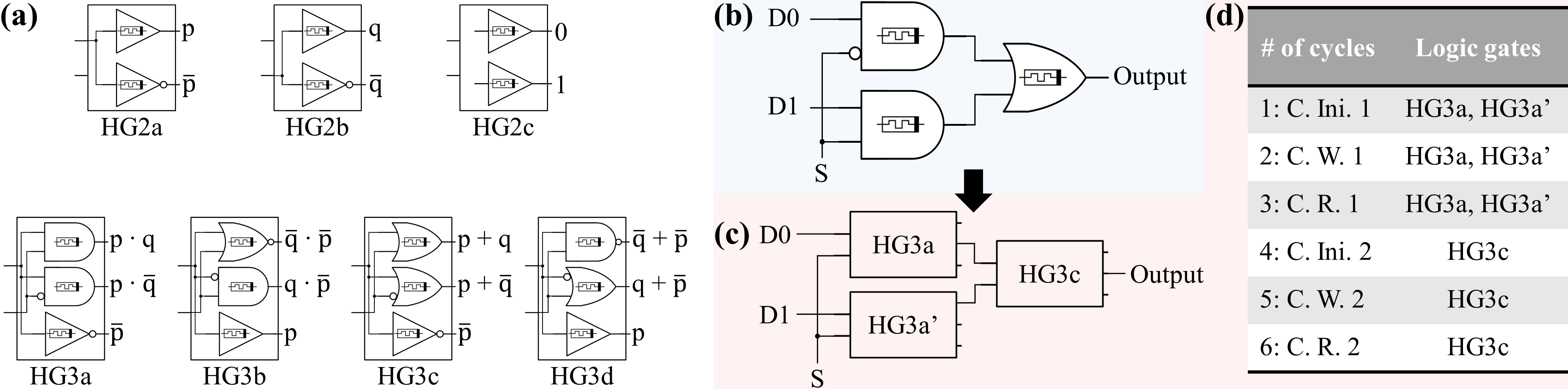}
  \caption{(a) Hiding groups consisting of 2-cell and 3-cell configurations. (b) Example of a 2-to-1 multiplexer circuit without the hiding strategy, implemented using three logic gates, with each gate requiring a memristor for implementation. (c) An equivalent 2-to-1 multiplexer achieved using hiding groups, effectively concealing power consumption by replacing individual gates with two HG3a hiding groups (HG3a and HG3a') and HG3c hiding group. (d) Cycle schedule for the implementation of the multiplexer based on the hiding groups.}
  \label{fig:MUX}
\end{figure*}

Fig. \ref{fig:14gates}a showcases the extensive range of gates achievable when initializing S' to 0, which can be achieved by applying 0 V to the TE and 6 V to the BE. Initializing the BiBFO memristor with an initialization state S' of '0' allows for the realization of gate types such as p $\cdot$ q, q $\cdot$ $\overline{\text{p}}$, p $\cdot$ $\overline{\text{q}}$, and $\overline{\text{q}}$ $\cdot$ $\overline{\text{p}}$. This advantage is derived from the 0-controlling property of the AND logic function, where 0 holds significance. Consequently, initializing the memristive cell to '0' simplifies the implementation of AND logic gates. Referring to the truth table in Fig. \ref{fig:IV}c under the S' = '0' initialization, the equation $\text{Out}^\text{w}$ = TE $\cdot$ $\overline{\text{BE}}$ succinctly summarizes the scenario, enabling the realization of gates achieved through the AND operation with two input variables, as depicted in Fig. \ref{fig:14gates}a.

In Fig. \ref{fig:14gates}b, a comprehensive demonstration of all possible gates under the initialization of S' = '1' is presented, accomplished by applying 6 V to the TE and 0 V to the BE. By utilizing the BiBFO memristor and initializing the memristive cell with an initialization state of '1', gate types including p $+$ q, q $+$ $\overline{\text{p}}$, p $+$ $\overline{\text{q}}$, and $\overline{\text{p}}$ $+$ $\overline{\text{q}}$ can be realized. This advantage arises from the 1-controlling nature of the OR logic function, with 1 being the most significant bit. Therefore, achieving OR logic gates becomes more accessible and advantageous when the initialization state is set to '1'. Referring to the truth table in Fig. \ref{fig:IV}c under the initialization of S' = '1', the equation $\text{Out}^\text{w}$ = TE $+$ $\overline{\text{BE}}$ succinctly summarizes the scenario, enabling the realization of gates calculated through the OR operation with two input variables, as illustrated in Fig. \ref{fig:14gates}b. Similarly, to achieve the gate q $+$ $\overline{\text{p}}$, the LIV1 = q is applied to the TE, while $\overline{\text{LIV2}}$ = p is applied to the BE, following the provided equation. 

\begin{figure*}[htb]
  \includegraphics[width=1\linewidth]{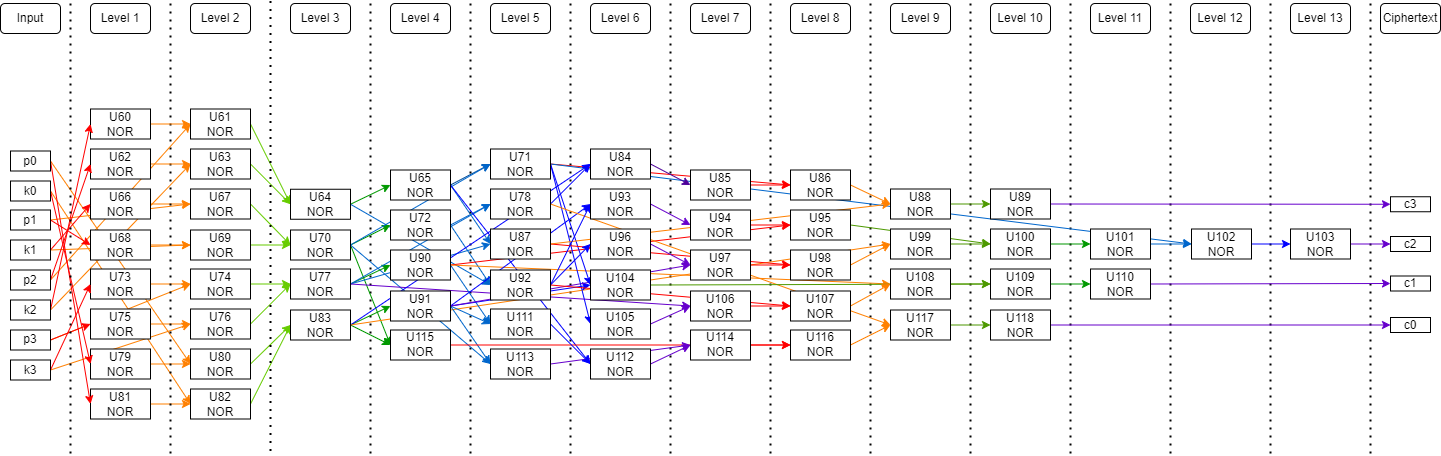}
  \caption{Synthesized gate-level netlist of xor4SBox based on NOR gates, with each individual NOR gate implemented using a BiBFO memristor.}
  \label{fig:Sbox}
\end{figure*}

Both tables presented in Fig. \ref{fig:14gates} exhibit a notable left-right symmetry along the diagonal, where the left side region is depicted in light gray and the right side region in dark gray. This symmetry arises from the intrinsic properties of the AND and OR gates showcased in Fig. \ref{fig:14gates}a and Fig. \ref{fig:14gates}b, respectively. Specifically, the commutative law allows for the swapping of inputs in both the AND and OR gates without affecting the computed outcome. Consequently, under the initialization of S' = '0' and S' = '1', the common gate types q, p, $\overline{\text{q}}$, and $\overline{\text{p}}$ can be easily realized by setting the two input variables to the same values, i.e., LIV1 = LIV2.

We propose a novel power-balanced hiding method utilizing memristor groups to effectively conceal power consumption, ensuring consistent power costs regardless of the LIV values during writing and reading cycles. Our method guarantees equal power costs for the memristor group, regardless of specific LIV settings.
The gate types depicted in Fig. \ref{fig:14gates}a can be categorized into two sets of 3-cell hiding groups: HG3a (p $\cdot$ q, p $\cdot$ $\overline{\text{q}}$, $\overline{\text{p}}$) and HG3b ($\overline{\text{q}}$ $\cdot$ $\overline{\text{p}}$, q $\cdot$ $\overline{\text{p}}$, p). Similarly, in Fig. \ref{fig:14gates}b, the gate types are also divided into two sets of 3-cell hiding groups: HG3c (p $+$ q, p $+$ $\overline{\text{q}}$, $\overline{\text{p}}$) and HG3d ($\overline{\text{p}}$ $+$ $\overline{\text{q}}$, q $+$ $\overline{\text{p}}$, p).
Furthermore, our hiding strategy extends to the 2-cell hiding groups (HG2), enabling effective power hiding among common gate types. By considering the power of the individual gate and the power of its negation gate, the power consumption of a common gate type can be concealed. For example, the gate types p and $\overline{\text{p}}$ are considered as a single gate type in both HG3b and HG3a. All the 2-cell and 3-cell hiding groups are depicted in Fig. \ref{fig:MUX}a.

In Fig. \ref{fig:14gates}, gates that are marked with the same color can be operated together using the same input data to conceal power, under the same initialization. For example, the gates $\overline{\text{q}}$ $\cdot$ $\overline{\text{p}}$, q $\cdot$ $\overline{\text{p}}$, and p can be simultaneously operated. This simultaneous operation ensures that the power consumed during the writing cycle is the sum of A, B, and C, and the power consumed during the reading cycle with r = 1 is equal to the sum of W, W, and Y. Importantly, the sum of power during the writing and reading cycles remains independent of the input variables LIV1 and LIV2.

It should be noted that Fig. \ref{fig:14gates} provides an overview of the power hiding strategy for 14 gates when the reading variable r is set to 1. However, gates based on the OR/AND logic function can also be realized under the initialization of S' = '0'/'1', but with r = 0 and reversed input variables LIV1 and LIV2. This is because, as shown in Fig. \ref{fig:IV}d, when r is set to 0, $\text{Out}^\text{r}$ = $\overline{\text{Out}^\text{w}}$. For example, under the initialization of S' = '1', when r is changed from 1 to 0, $\overline{\text{q}}$ $+$ $\overline{\text{p}}$ will be transformed into $\overline{\overline{\text{q}} + \overline{\text{p}}}$, which is equivalent to p $\cdot$ q. This phenomenon highlights the significant influence of the reading variable, specifically the reading bias applied to the TE, on the realization of gate types. As shown in Fig. \ref{fig:14gates}, by solely utilizing writing cycles and selecting the reading variable as either 1 or 0, only 14 out of the 16 gate types can be achieved. The XOR and XNOR are the two missing gate types, which require a more intricate combination of writing cycles and a selection of reading variables involving q, p, $\overline{\text{q}}$, and $\overline{\text{p}}$ to be realized.

Fig. \ref{fig:XOR} illustrates all possible configurations for realizing XOR and XNOR gate types using BiBFO memristors by incorporating a reading cycle with a more complex selection of reading variables following the writing cycle.
In Fig. \ref{fig:XOR}a, the outputs of the writing cycle, denoted as $\text{Out}^\text{w}$, are set to p and $\overline{\text{p}}$. These outputs are then read out using the reading variables q and $\overline{\text{q}}$. The resulting outputs of the reading cycle, $\text{Out}^\text{r}$, are highlighted in bold, and the power marks associated with them are shown in red. The $\text{Out}^\text{r}$ values with a dark or light blue background indicate the successful implementation of XNOR or XOR gate types, respectively. Furthermore, these implemented XNOR and XOR can be combined to form a hiding group, allowing them to conceal power from each other. 

As a result, the sum of the power in the writing cycle can be either B + B + C + C or D + D + E + E. The two possibilities for the sum of the writing cycle power arise because the $\text{Out}^\text{w}$ as p and $\overline{\text{p}}$ can be achieved with the initialization of S' as '0' or '1' and r = 1, as demonstrated in Fig. \ref{fig:14gates}a and Fig. \ref{fig:14gates}b. Simultaneously, the sum of the power in the reading cycle can be observed from Fig. \ref{fig:XOR}a, which is equal to W + X + Y + Z. In Fig. \ref{fig:XOR}b, another combination of XOR and XNOR is presented as a hiding group, utilizing the $\text{Out}^\text{w}$ outputs q and $\overline{\text{q}}$ with reading variables p and $\overline{\text{p}}$. The power consumption of this hiding group, including two XNOR (dark yellow color) and two XOR (light yellow color), in the writing cycle (B + B + C + C or D + D + E + E) and reading cycle (W + X + Y + Z) is effectively concealed.

To elucidate the functioning of our proposed hiding groups in concealing power consumption of circuit during operation, we chose to construct an example circuit in the form of a 2-to-1 multiplexer. Fig. \ref{fig:MUX}b illustrates such multiplexer circuit without the hiding strategy built using three logic gates (specifically, p $\cdot$ $\overline{\text{q}}$, p $\cdot$ q, and p $+$ q), with each gate requiring a BiBFO memristor for implementation. In the 2-to-1 multiplexer circuit, denoted as D0 and D1 for the two inputs and S for the selection signal, the output corresponds to D0 or D1 based on the value of S (0 or 1).

In order to demonstrate the operational principle of our proposed hiding groups, an equivalent 2-to-1 multiplexer is constructed using hiding groups, as illustrated in Fig. \ref{fig:MUX}c. By utilizing two HG3a groups (HG3a and HG3a') in place of the individual p $\cdot$ $\overline{\text{q}}$ and p $\cdot$ q gates and HG3c to replace the p $+$ q gate, the power consumption of the multiplexer circuit is effectively concealed, due to the balanced power consumption within each hiding group in each cycle. The implementation of this 2-to-1 multiplexer based on hiding groups necessitates a total of 6 cycles, as indicated in Fig. \ref{fig:MUX}d. The two parallel HG3a and HG3a' groups collectively occupy 3 cycles, specifically C. Ini. 1, C. W. 1, and C. R. 1. Similarly, HG3c occupies 3 cycles, denoted as C. Ini. 2, C. W. 2, and C. R. 2.

\begin{figure*}[htb]
\centering
  \includegraphics[width=\linewidth]{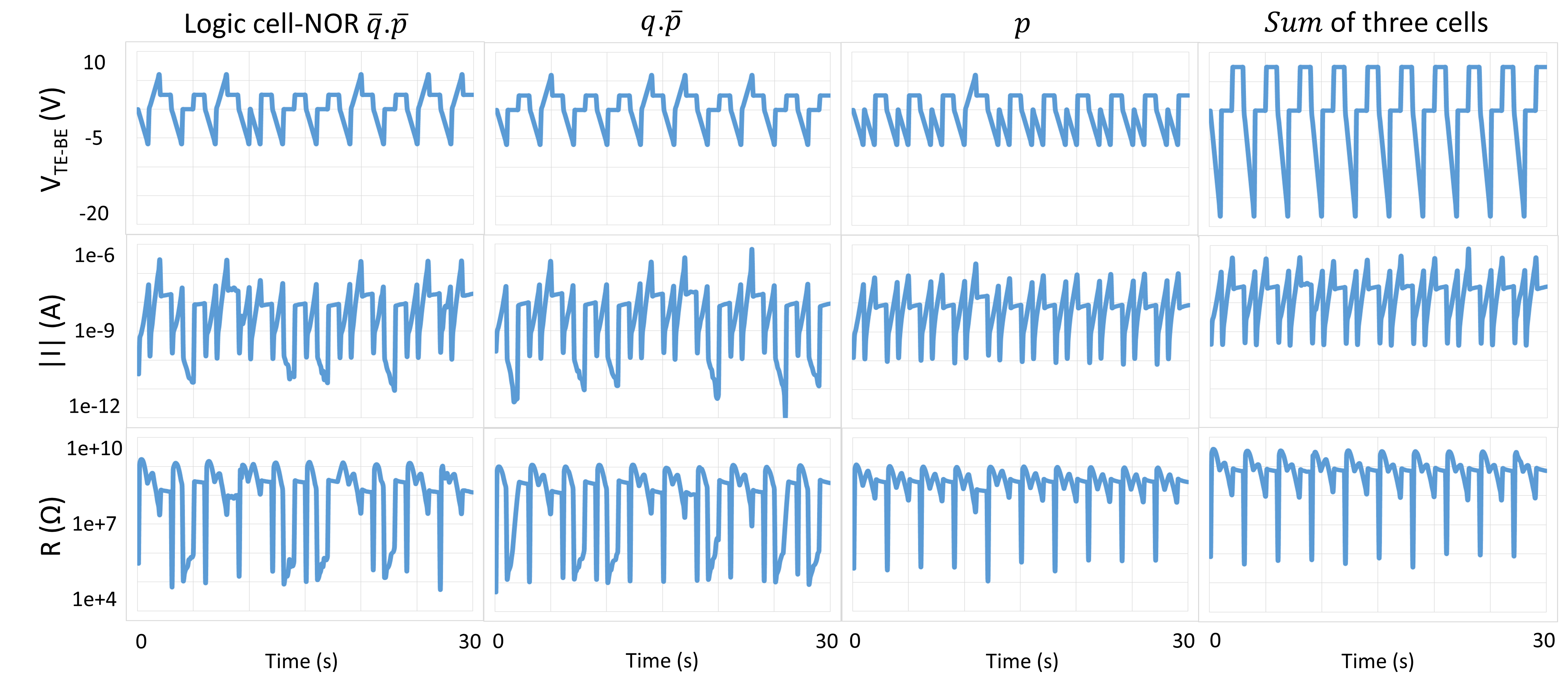}
  \caption{Experimental electrical execution of the memristor M1 ($\overline{\text{q}}$ $\cdot$ $\overline{\text{p}}$) and other two memristors (q $\cdot$ $\overline{\text{p}}$ and p) that belong to its hiding group HG3b to realize the xor4SBox with plaintext $= 0000$ with key $= 0000$ under the hiding strategy. The rightmost column of data presents the overall experimental results for hiding group HG3b.}
  \label{fig:waveform}
\end{figure*}

\begin{table}[htb]
\begin{center}
\caption{Implementation tables for the NOR gate and other two gates within its hiding group HG3b, depicting power marks in the writing and reading cycles.}
\label{tab1:NOR}
\begin{tabular}{ccccclccl}
\hline
\multicolumn{3}{c|}{}                                                    & \multicolumn{2}{c|}{C. Ini.}                          & \multicolumn{2}{c|}{C. W.}                        & \multicolumn{2}{c}{C. R.}       \\ \cline{4-9} 
\multicolumn{3}{c|}{\multirow{-2}{*}{$\overline{\text{q}}$ $\cdot$ $\overline{\text{p}}$ (NOR)}}                               & \multicolumn{1}{c|}{TE}   & \multicolumn{1}{c|}{BE}   & \multicolumn{1}{c|}{TE} & \multicolumn{1}{c|}{BE} & \multicolumn{2}{c}{r}           \\ \hline
\multicolumn{1}{c|}{q} & \multicolumn{1}{c|}{p} & \multicolumn{1}{c|}{s} & \multicolumn{1}{c|}{0}    & \multicolumn{1}{c|}{1}    & \multicolumn{1}{c|}{$\overline{\text{q}}$} & \multicolumn{1}{c|}{p}  & \multicolumn{2}{c}{1}           \\ \hline
0                      & 0                      & \multicolumn{1}{c|}{1} & \multicolumn{2}{c|}{\{PH,NL\}} & \multicolumn{2}{l|}{\{PL,NH\}  \color{red}C}                 & \multicolumn{2}{c}{PL =:’1’  \color{red}Y} \\
1                      & 0                      & \multicolumn{1}{c|}{0} & \multicolumn{2}{c|}{\{PH,NL\}}                        & \multicolumn{2}{l|}{\{PH,NL\}  \color{red}A}                 & \multicolumn{2}{c}{PH =:’0’  \color{red}W} \\
0                      & 1                      & \multicolumn{1}{c|}{0} & \multicolumn{2}{c|}{\{PH,NL\}}                        & \multicolumn{2}{l|}{\{PH,NL\}  \color{red}A}                 & \multicolumn{2}{c}{PH =:’0’  \color{red}W} \\
1                      & 1                      & \multicolumn{1}{c|}{0} & \multicolumn{2}{c|}{\{PH,NL\}}                        & \multicolumn{2}{l|}{\{PH,NL\}  \color{red}B}                 & \multicolumn{2}{c}{PH =:’0’  \color{red}W} \\ \hline
\multicolumn{1}{l}{}   & \multicolumn{1}{l}{}   & \multicolumn{1}{l}{}   & \multicolumn{1}{l}{}      & \multicolumn{1}{l}{}      &                         & \multicolumn{1}{l}{}    & \multicolumn{1}{l}{}     &      \\ \hline
\multicolumn{3}{c|}{}                                                    & \multicolumn{2}{c|}{C. Ini.}                          & \multicolumn{2}{c|}{C. W.}                        & \multicolumn{2}{c}{C. R.}       \\ \cline{4-9} 
\multicolumn{3}{c|}{\multirow{-2}{*}{q $\cdot$ $\overline{\text{p}}$}}                                & \multicolumn{1}{c|}{TE}   & \multicolumn{1}{c|}{BE}   & \multicolumn{1}{c|}{TE} & \multicolumn{1}{c|}{BE} & \multicolumn{2}{c}{r}           \\ \hline
\multicolumn{1}{c|}{q} & \multicolumn{1}{c|}{p} & \multicolumn{1}{c|}{s} & \multicolumn{1}{c|}{0}    & \multicolumn{1}{c|}{1}    & \multicolumn{1}{c|}{q}  & \multicolumn{1}{c|}{p}  & \multicolumn{2}{c}{1}           \\ \hline
0                      & 0                      & \multicolumn{1}{c|}{0} & \multicolumn{2}{c|}{\{PH,NL\}}                        & \multicolumn{2}{l|}{\{PH,NL\}  \color{red}A}                 & \multicolumn{2}{c}{PH =:’0’  \color{red}W} \\
1                      & 0                      & \multicolumn{1}{c|}{1} & \multicolumn{2}{c|}{\{PH,NL\}}                        & \multicolumn{2}{l|}{\{PL,NH\}  \color{red}C}                 & \multicolumn{2}{c}{PL =:’1’  \color{red}Y} \\
0                      & 1                      & \multicolumn{1}{c|}{0} & \multicolumn{2}{c|}{\{PH,NL\}}                        & \multicolumn{2}{l|}{\{PH,NL\}  \color{red}B}                 & \multicolumn{2}{c}{PH =:’0’  \color{red}W} \\
1                      & 1                      & \multicolumn{1}{c|}{0} & \multicolumn{2}{c|}{\{PH,NL\}}                        & \multicolumn{2}{l|}{\{PH,NL\}  \color{red}A}                 & \multicolumn{2}{c}{PH =:’0’  \color{red}W} \\ \hline
\multicolumn{1}{l}{}   & \multicolumn{1}{l}{}   & \multicolumn{1}{l}{}   & \multicolumn{1}{l}{}      & \multicolumn{1}{l}{}      &                         & \multicolumn{1}{l}{}    & \multicolumn{1}{l}{}     &      \\ \hline
\multicolumn{3}{c|}{}                                                    & \multicolumn{2}{c|}{C. Ini.}                          & \multicolumn{2}{c|}{C. W.}                        & \multicolumn{2}{c}{C. R.}       \\ \cline{4-9} 
\multicolumn{3}{c|}{\multirow{-2}{*}{p}}                                & \multicolumn{1}{c|}{TE}   & \multicolumn{1}{c|}{BE}   & \multicolumn{1}{c|}{TE} & \multicolumn{1}{c|}{BE} & \multicolumn{2}{c}{r}           \\ \hline
\multicolumn{1}{c|}{q} & \multicolumn{1}{c|}{p} & \multicolumn{1}{c|}{s} & \multicolumn{1}{c|}{0}    & \multicolumn{1}{c|}{1}    & \multicolumn{1}{c|}{p}  & \multicolumn{1}{c|}{$\overline{\text{p}}$} & \multicolumn{2}{c}{1}           \\ \hline
0                      & 0                      & \multicolumn{1}{c|}{0} & \multicolumn{2}{c|}{\{PH,NL\}}                        & \multicolumn{2}{l|}{\{PH,NL\}  \color{red}B}                 & \multicolumn{2}{c}{PH =:’0’  \color{red}W} \\
1                      & 0                      & \multicolumn{1}{c|}{0} & \multicolumn{2}{c|}{\{PH,NL\}}                        & \multicolumn{2}{l|}{\{PH,NL\}  \color{red}B}                 & \multicolumn{2}{c}{PH =:’0’  \color{red}W} \\
0                      & 1                      & \multicolumn{1}{c|}{1} & \multicolumn{2}{c|}{\{PH,NL\}}                        & \multicolumn{2}{l|}{\{PL,NH\}  \color{red}C}                 & \multicolumn{2}{c}{PL =:’1’  \color{red}Y} \\
1                      & 1                      & \multicolumn{1}{c|}{1} & \multicolumn{2}{c|}{\{PH,NL\}}                        & \multicolumn{2}{l|}{\{PL,NH\}  \color{red}C}                 & \multicolumn{2}{c}{PL =:’1’  \color{red}Y} \\ \hline
\end{tabular}
\end{center}
\end{table}

\section{Experimental validation}
\label{sec:Valid.}

In Table \ref{tab1:NOR}, the implementation table for the NOR gate and its hiding group HG3b is presented as an exemplary case to experimentally validate the power-balanced hiding strategy. These gates, including the NOR gate, share a common initialization condition of S' = '0', achieved by applying a bias of 0 V to the TE and 6 V to the BE. The three gate types highlighted in red in Fig. \ref{fig:14gates}a can be effectively realized through a combination of writing and reading cycles with r = 1. Importantly, the total power consumption of this hiding group, represented by the sum A + B + C in the writing cycle and W + W + Y in the reading cycle, remains independent of the input variables LIV1 and LIV2.

To validate the effectiveness of the hiding strategy, a cryptographic construction known as xor4SBox based on NOR gates has been employed. This construction utilizes a 4-bit "cipher" and operates on the plaintext ($p_3$, $p_2$, $p_1$, $p_0$), the secret key ($k_3$, $k_2$, $k_1$, $k_0$), and generates the corresponding ciphertext ($c_3$, $c_2$, $c_1$, $c_0$) using the equation $S(p_3 \oplus k_3$, $p_2 \oplus k_2$, $p_1 \oplus k_1$, $p_0 \oplus k_0$), i.e. small-scale version of the AES S-Box \cite{CMR:05}. By leveraging the hiding strategy outlined in Fig. \ref{fig:14gates} and Fig. \ref{fig:XOR}, the feasibility of the approach in ensuring secure and efficient cryptographic operations is demonstrated. Note that, the efficacy of the proposed hiding strategy is apparent in idealized memristor based circuits devoid of any C2C and D2D variations. However, practical circuits, subject to inherent variability and minor inconsistencies, emphasize the necessity for experimental measurements and rigorous non-specific T-tests \cite{GJJR:11} to ensure a comprehensive evaluation of the effectiveness of the hiding strategy in xor4SBox.

Fig. \ref{fig:Sbox} depicts the construction achieved using a synthesis tool based on NOR logic gates. The circuit consists of 59 NOR gates arranged in 13 levels and is implemented on N = 8 memristors (M1-M8) without considering the hiding strategy. However, by incorporating the hiding strategy, an additional 2N = 16 gates are required, resulting in a total of 3N = 24 memristors being utilized to realize the xor4SBox based on NOR gates.

\begin{figure}[htb]
  \includegraphics[width=\linewidth]{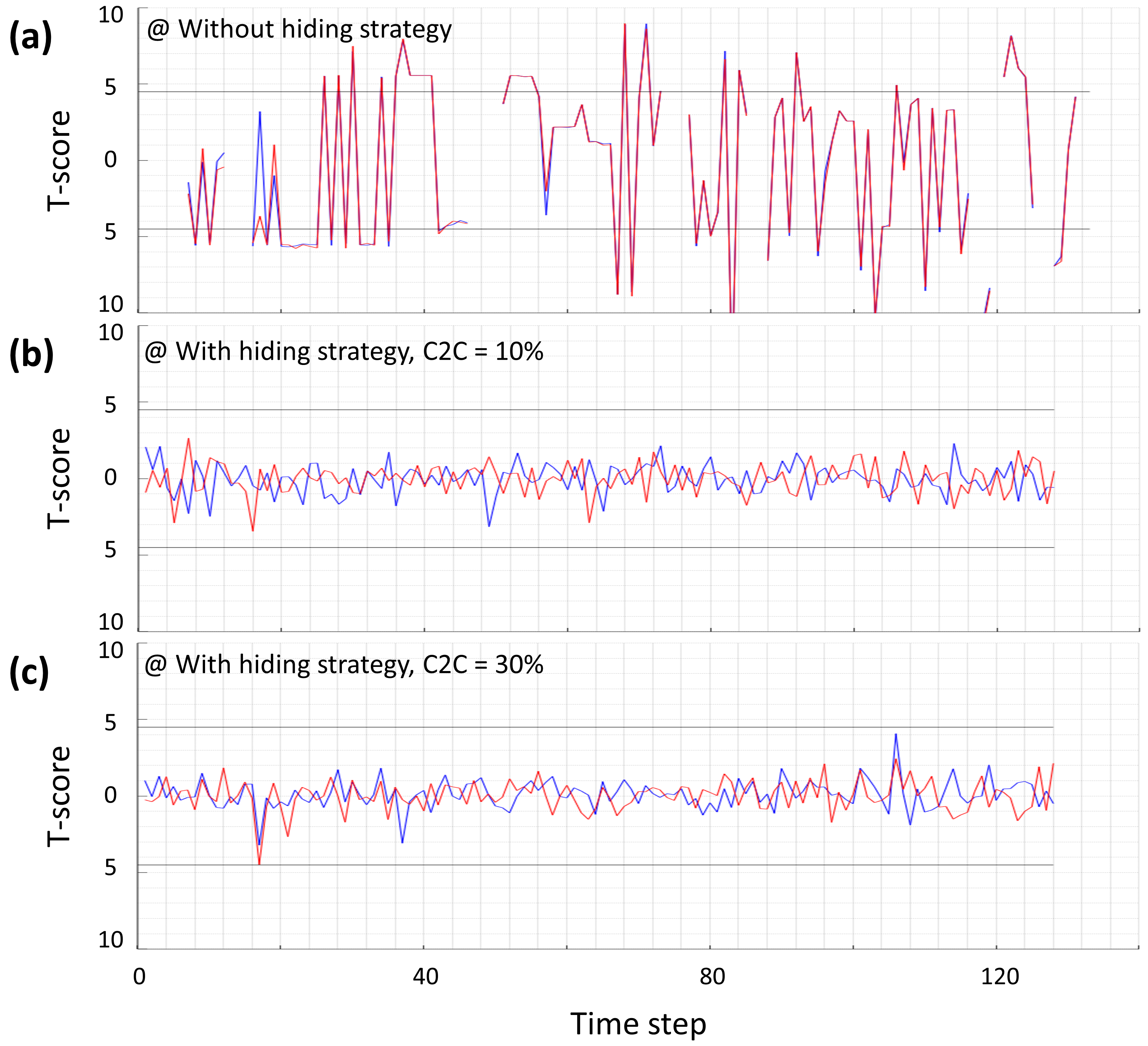}
  \caption{T-test diagrams of xor4SBox based on NOR gates (a) without hiding strategy and with hiding strategy at C2C variations of (b) 10\% and (c) 30\%.}
  \label{fig:t-test}
\end{figure}

In this work, we have employed BiBFO memristors with uniform TE dimensions and a shared BE configuration, enabling their utilization as a line array. To conduct electrical measurements on these BiBFO memristive line arrays, we utilized a Keithley sourcemeter 2400 in conjunction with a Formfactor EPS150 Probe station and an 8 by 1 probe card featuring a 600 nm pitch.
For experimental testing involving the small-scale memristive S-Box in Fig. \ref{fig:Sbox} with considering the hiding strategy, a total of 3 memristive line arrays are operated according to a sequential scheme, with each memristive line array comprising 8 BiBFO memristors. The requisite voltage biases for initialization, writing, and reading operations are generated by a sourcemeter under the precise control of a synthesis-LABVIEW program through the switching matrix fabricated on a PCB board.

Specifically, memristor M1 is employed iteratively across multiple levels from Level 1 to Level 13, where it serves to implement the first NOR gate within each level, denoted as U60, U61, U64, U65, U71, U84, U85, U86, U88, U89, U101, U102, U103. Similarly, M2 is dedicated to realizing the second NOR gate in each level, followed by the subsequent utilization of M3 through M8 to complete the remaining NOR gates in the multiple levels.
Fig. \ref{fig:waveform} presents the experimental results of memristor M1 and other two memristors that belong to its hiding group HG3b in the implementation of xor4SBox using BiBFO memristors. The experiment involved a key of 0000 and a plaintext of 0000, with measurements including experimental resistance values, voltage difference between the BE and TE, and the absolute values of current across each cell $\lvert I \rvert$. The rightmost column of data illustrates the comprehensive experimental outcomes for hiding group HG3b.

As depicted in Fig. \ref{fig:waveform}, the experimental results of the three individual memristors within HG3b exhibit irregular variations over time (i.e., as cycles progress). This irregular behavior poses a risk of data leakage from the ongoing computations. Nevertheless, the overall experimental results for hiding group HG3b remain relatively consistent throughout each cycle, effectively concealing the data being processed.
Note that arbitrary writing bias waveforms of $+V_\text{w}$/$-V_\text{w}$ for 100 ms can set/reset BiBFO memristors. In this work, triangular biases are chosen for the initialization and writing cycles during the measurement process to enhance device endurance. Nevertheless, the reading bias $V_\text{r}$ = 2.5 V, characterized by its small amplitude, persists in its pulsed waveform, as illustrated in Fig. \ref{fig:waveform}.

To assess the existence of information leakage of the xor4SBox implementation, T-tests are employed both in the absence and presence of the power-balanced hiding technique. To address the extensive data requirements necessitated by the T-test, our prior work \cite{chen2022side} saw the development of a simulator tailored for conducting memristive cryptographic circuits, grounded in the memristive model. In order to align simulation outcomes more closely with authentic physical measurements, we incorporated adjustable parameters for the C2C values within the simulator.
In this work, we utilize this simulator to conduct the xor4SBox with and without considering hiding strategy, under a time resolution of one step per millisecond. This simulator interprets Boolean values (0 and 1) as its inputs, transmuting them into triangular pulses at the device's TE and BE, which corresponds to the operation of our memristive experimental setup. Subsequently, the resultant 64 power traces are meticulously subjected to analysis using Octave to compute the T-score.

A comparison between the T-test results of xor4SBox without the hiding strategy (Fig. \ref{fig:t-test}a) and with the hiding strategy, featuring C2C values of 10\% (Fig. \ref{fig:t-test}b) and 30\% (Fig. \ref{fig:t-test}c), reveals significant differences. In our study, we investigated BFO-based memristive devices, characterized by their analog interfacial switching mechanism. Under the specific fabrication conditions outlined in the paper, we anticipate a maximum cycle-to-cycle (C2C) variability of 30\%. This variability is inherent to the analog nature of interfacial switching in these devices.
In contrast, memristors with abrupt switching behaviors exhibit potentially larger device variations due to their filamentary switching nature.

The T-test conducted on xor4SBox without the hiding strategy yields unsuccessful results, indicating the presence of information leakage during its operation. The discontinuity observed in the T-test line depicted in Fig. \ref{fig:t-test}a is attributed to one of the traces experiencing a momentary absence of power, essentially remaining in a floating state during that particular interval. On the other hand, the xor4SBox implemented with the hiding strategy, regardless of the C2C value at 10\% or 30\%, successfully passes the T-test, remaining within the upper and lower boundaries. These findings provide compelling evidence for the feasibility and effectiveness of our proposed hiding strategy.

\section{Conclusion}
\label{sec:Conclu.}

In conclusion, we have proposed a novel power-balanced hiding strategy utilizing memristor hiding groups to effectively conceal power consumption in logic circuits. Our strategy ensures consistent power costs regardless of the LIV values during writing and reading cycles. By categorizing gate types into hiding groups and employing simultaneous operations, we achieved power hiding among all 16 logic gate types. For 14 out of the 16 logic gate types (excluding XOR and XNOR), the realization of power hiding is accomplished by employing BiBFO memristors with a single writing cycle, and setting the reading variable r to 1. In the case of XOR and XNOR, the attainment of power hiding is achieved through intricate combinations of writing and reading cycles, highlighting the influence of the reading variable on gate types.

Through experimental validation, we showcased the effectiveness of our hiding strategy in implementing a cryptographic construction and ensuring secure and efficient cryptographic operations. The construction of the xor4SBox circuit employing NOR gates, with and without the hiding strategy, was presented, and the T-test analysis confirmed the successful elimination of information leakage with our proposed hiding strategy. These findings present a significant advancement in power-balanced hiding methods for enhancing the security and efficiency of cryptographic logic circuits by employing memristive technology.

\section*{Acknowledgments}
This work is supported by the German Research Foundation (DFG) Project MemCrypto (Grant Nr. DU 1896/2-1 and PO 1220/15-1) in SPP 2253 Nano Security. 

\bibliographystyle{IEEEtran}
\bibliography{Main}


\section{Biography Section}
\begin{IEEEbiography}[{\includegraphics[width=1in,height=1.25in,clip,keepaspectratio]{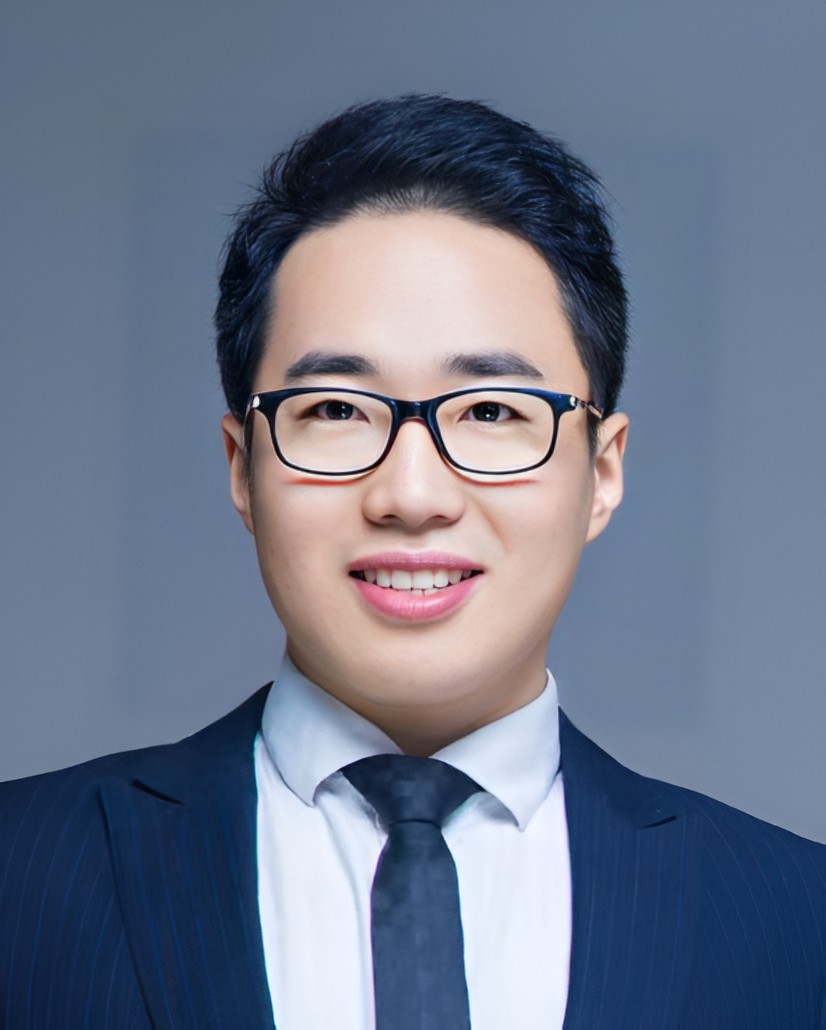}}]{Ziang Chen}received his bachelor's degree in microelectronics manufacturing engineering from the Guilin University of Electronic Technology, China, in 2014, and his M.S. degree in microsystems and microelectronics from the Technical University of Chemnitz, Chemnitz, Germany, in 2019. He is currently pursuing the Ph.D. degree with the Institute for Solid State Physics, Friedrich Schiller University Jena, Jena, Germany. His research interests include memristive devices based crossbar topology, hardware security and neuromorphic computing exploiting memristive devices.
\end{IEEEbiography}

\begin{IEEEbiography}[{\includegraphics[width=1in,height=1.25in,clip,keepaspectratio]{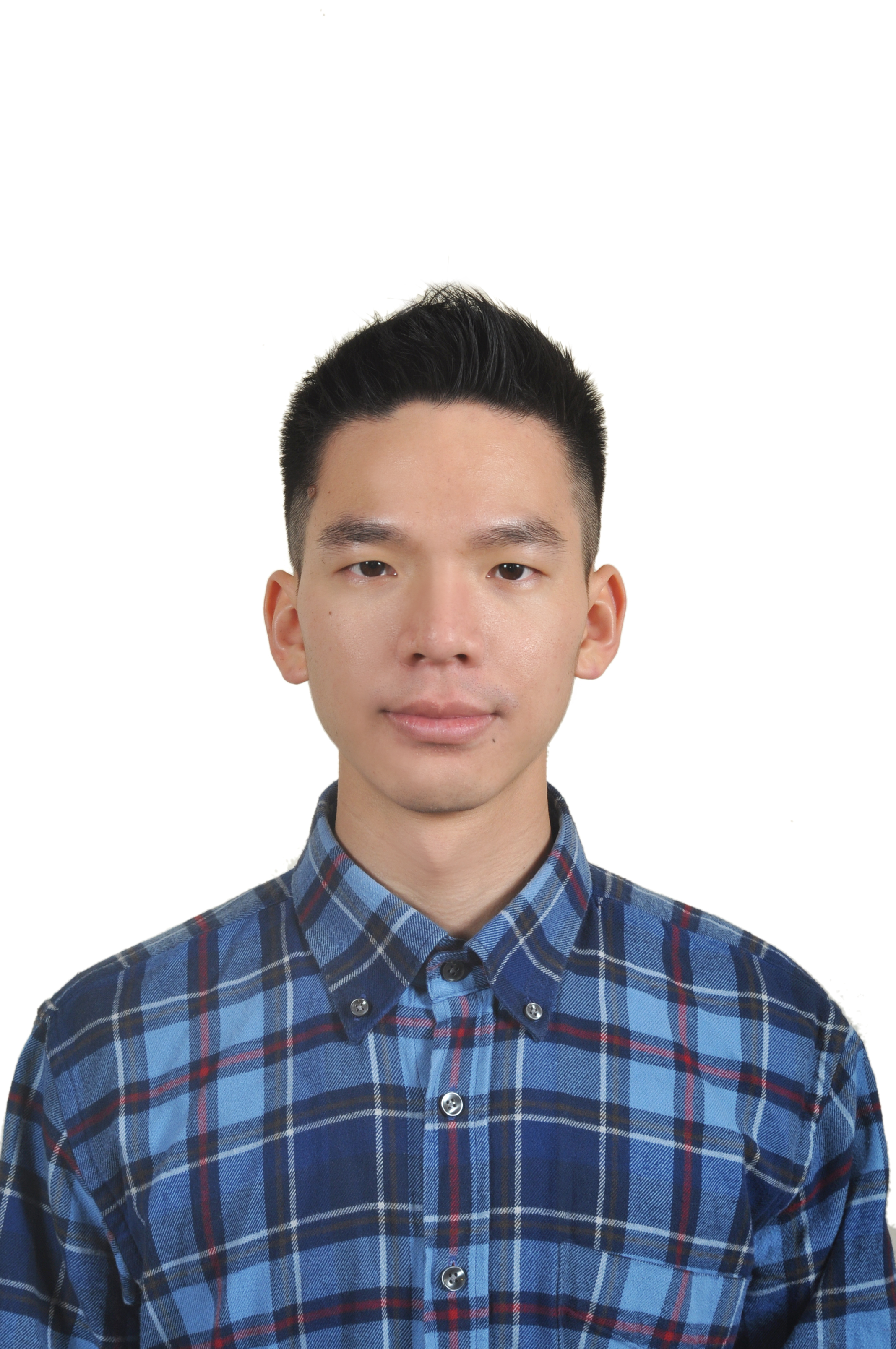}}]{Li-Wei Chen}received his Bachelor of Science (B.Sc.) degree in Electrical Engineering from Georgia Institute of Technology, Atlanta, USA. Afterward, he continued his studies at Universität Stuttgart, Germany; and received his Master of Science (M.Sc.) degree in Information Technology in 2020. He joined the Institute of Computer Architecture and Computer Engineering at Universität Stuttgart as a research assistant in October 2020. His current research interests include cryptography, hardware-oriented security, and memristor.
\end{IEEEbiography}

\begin{IEEEbiography}[{\includegraphics[width=1in,height=1.25in,clip,keepaspectratio]{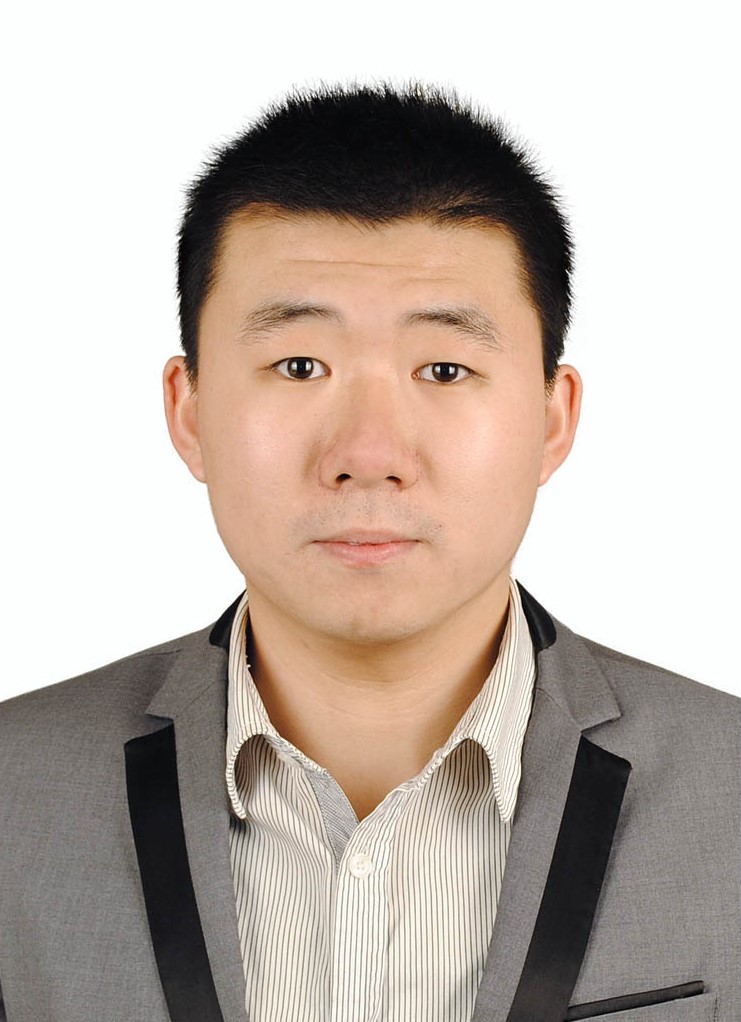}}]{Xianyue Zhao}received the B.S. degree in electrotechnical science and technology from Yanshan University, Qinhuangdao, China, in 2013 and the M.S. degree in microelectronics from Chemnitz University of Technology, Chemnitz, Germany, in 2019. He is currently pursuing the Ph.D. degree with the Institute for Solid State Physics, Friedrich Schiller University Jena, Jena, Germany. His research interests include memristive devices in switching mechanisms, wafer-level memristive components, and memristive device based hardware-oriented security.
\end{IEEEbiography}

\begin{IEEEbiography}[{\includegraphics[width=1in,height=1.25in,clip,keepaspectratio]{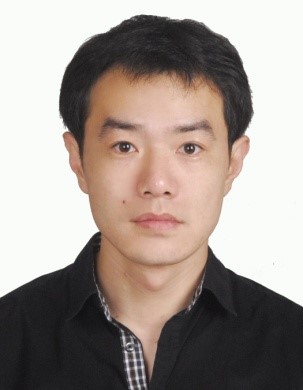}}]{Kefeng Li}received Master degree in Condensed matter physics in 2008, and received the Ph.D. degree in Metallurgical Engineering in 2014. He currently works as a postdoctor at Friedrich Schiller University Jena. 2015-2017 he was a postdoctor at Technische Universität Chemnitz, then he was a senior materials R\&D engineer at Guangdong Academy of Sciences, China. His research interests are the materials design and microstructure characterization.
\end{IEEEbiography}

\begin{IEEEbiography}[{\includegraphics[width=1in,height=1.25in,clip,keepaspectratio]{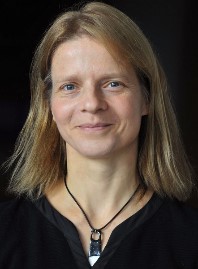}}]{Prof. Dr. Heidemarie Schmidt}
studied Physics at the Technical University Leipzig where she graduated in 1999 with a dissertation on band structures in ultrathin semiconductors. At the Institute for Experimental Physics II of the University of Leipzig, she headed the BMBF junior research group "Nano-Spintronics" from 2003 to 2007. From 2007 she established a junior research group of the same name at the Institute for Ion Beam Physics and Materials Research in the Helmholtz Center Dresden-Rossendorf. In 2012, Schmidt received a Heisenberg scholarship from the Deutsche Forschungsgemeinschaft (DFG) and started to work on electroforming-free hardware materials for AI, namely {$\rm BiFeO_{3}$} and {$\rm YMnO_{3}$} (Chemnitz University of Technology, Faculty of Electrical Engineering and Information Technology). Since 2016, she and her BFO4ICT-ATTRACT group develop the technology for an industrial production of BFO memristors at the Fraunhofer Institute ENAS in Chemnitz. In September 2017 she became Professor with focus on quantum detection at the Institute of Solid State Physics of the Friedrich Schiller University Jena and took over the management of the Quantum Detection Department at the Leibniz Institute for Photonic Technology Jena.
\end{IEEEbiography}

\begin{IEEEbiography}[{\includegraphics[width=1in,height=1.25in,clip,keepaspectratio]{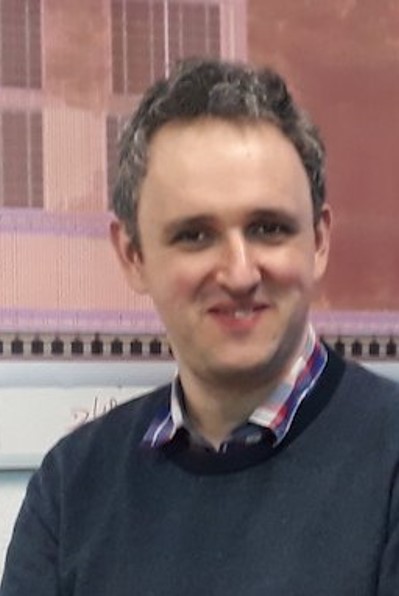}}]{Ilia Polian}is a Full Professor and the Director of the Institute for Computer Architecture and Computer Engineering at the University of Stuttgart, Germany. He received his Diplom (MSc) and PhD degrees from the University of Freiburg, Germany, in 1999 and 2003, respectively. Prof. Polian co-authored over 200 scientific publications and received two Best Paper Awards. He is a Senior Member of IEEE. His scientific interest include hardware-oriented security, emerging architectures, test methods, and quantum computing.
\end{IEEEbiography}

\begin{IEEEbiography}[{\includegraphics[width=1in,height=1.25in,clip,keepaspectratio]{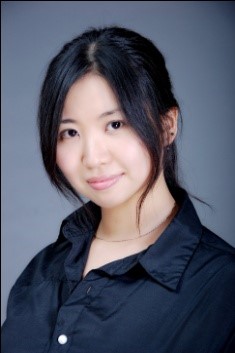}}]{Nan Du}received Diploma degree in Electrical Engineering at Faculty of Electrical Engineering and Information Technology at the Technical University of Dresden, Germany in 2011, and received the Ph.D. degree in Electrical Engineering at Technical University of Chemnitz, Germany, in 2016.

Since 2008 she is a member of the scientific staff of Prof. Dr. Heidemarie Schmidt. Between 2017 and 2019 she works as Postdoc at Fraunhofer ENAS in Chemnitz. In October 2019 she became Group Leader for research group “MemDevice” at Friedrich-Schiller-Universität Jena, Jena, Germany and at Fraunhofer ENAS, Chemnitz, Germany. She is corresponding author/coauthor of more than 30 publications and 12 inventions (information from depatisnet.dpma.de). Her research interests include the development of innovative security-oriented primitives and unconventional computing systems exploiting resistive switching devices, i.e., brain-inspired neuromorphic computing, in-memory computing and probabilistic computing.
Dr. Nan Du is member of coordination board of priority program “NanoSecurity” in Deutsch Forschungsgemeinschaft since 2020.
\end{IEEEbiography}

\vfill

\end{document}